# Observation of a first order phase transition to metal hydrogen near 425 GPa.

*Paul Loubeyre[1], Florent Occelli[1] and Paul Dumas[1&2].*

1 CEA, DAM, DIF, 91297 Arpajon. France.

2 Synchrotron SOLEI, 91192 Gif-sur-Yvette. France.

**Hydrogen has been the essential element in the development of atomic and molecular physics[1]. Moving to the properties of dense hydrogen has appeared a good deal more complex than originally thought by Wigner and Hungtinton in their seminal paper predicting metal hydrogen[2]: the electrons and the protons are strongly coupled to each other and ultimately must be treated equally[3,4]. The determination of how and when molecular solid hydrogen will transform into a metal is the stepping stone towards a full understanding of the quantum-many body properties of dense hydrogen. The quest for metal hydrogen has pushed major developments of modern experimental high pressure physics, yet the various claims of its observation over the past 30 years have remained controversial[5,6,7]. Here we show a first order phase transition near 425 GPa from insulator molecular solid hydrogen to metal hydrogen. Pressure in excess of 400 GPa could be achieved by using the recently developed Toroidal Diamond Anvil Cell (T-DAC)[8]. The structural and electronic properties of dense solid hydrogen at 80 K have been characterized by synchrotron infrared spectroscopy. The continuous vibron frequency shift and the electronic band gap closure down to 0.5 eV, both linearly evolving with pressure, point to the stability of the insulator C2/c-24 phase up to the metallic transition. Upon pressure release, the metallic state transforms back to the C2/c-24 phase with almost no hysteresis, hence suggesting that the metallization proceeds through a structural transformation within the molecular solid, presumably to the Cmca-12 structure. Our results are in good agreement with the scenario recently disclosed by an advanced calculation able to capture many-body electronic correlations[9].**

The search of metal hydrogen has a unique place in high pressure physics. Indisputably, metal hydrogen should exist. Due to increase in electrons kinetic energy because of quantum confinement, pressure should turn any insulator into a metal, as observed for molecular oxygen around 100 GPa some 20 years ago[10]. The early assumption was that the insulator-metal transition in hydrogen would be associated to the molecular dissociation[2]. However, it was later suggested that metal hydrogen may exist as a proton paired metal[11]. Quantitative predictions of the stability field and the properties of metal Hydrogen remain challenging because many contributions could be operational and should be self-consistently treated[3,4]: many-body electronic correlations, nuclear quantum effects (NQEs), nuclear spin ordering, coupling between protons and electrons as suggested by a large Born-Oppenheimer separation parameter, anharmonic effects, ... The most advanced calculations are now going beyond the electronic correlation mean-field description of density functional theory and try to capture many-body electronic correlations, such as the Diffusion Monte-Carlo simulation (DMC)[4,9,12].

But hydrogen is more than a model problem in condensed matter physics. Metal hydrogen is the ultimate hydride. It may exhibit a room temperature superconductivity[13,14,15], a melting transition at very low temperature into an unusual superconducting-superfluid state[16], a high protonic diffusion[17] and a high energy density storage. In 2004, Neil Ashcroft conjectured that some of the properties of metal hydrogen could be obtained at much lower pressure in hydrogen rich-systems[18]. This has led to the discovery of a novel class of materials, the super-hydrides, associated to recent measurements of record superconductivity temperatures in $H3S$[19] and $LaH10$[20], respectively 200K and 260 K. Also, metal hydrogen in a dense fluid state is central in planetary interiors. Investigating the quantum many body effects in solid metal hydrogen at low temperature could force physicists to rethink their model of hydrogen at deep planetary conditions where quantum effects prevail even at high temperature due to extreme density.

Previously, we measured the closure of the direct electronic bandgap of solid hydrogen in the visible up to above 300 GPa[21]. By extrapolating to zero the linear decrease of the electronic bandgap with pressure, the location of the insulator-metal transition in hydrogen was then predicted to occur around 450 GPa. In the present work, we follow the electronic bandgap closure in the near-to-mid infrared energy regions. IR measurements provide a non-invasive approach to characterize both the structural and the electronic properties of hydrogen up to its metallic state. Our experimental approach is based on two developments. First, to overcome the 400 GPa limit of conventional DAC[22], we used the novel T-DAC that enables the generation of pressure up to at least 600 GPa[8]. Importantly, it preserves the standard DAC quality of the measurements, in terms of stress distribution, optical access and sample size. Synthetic IIas diamonds were used to provide IR transparency down to 800 $cm^{-1}$. Second, we designed an infrared horizontal microscope which is coupled to a collimated exit port of a synchrotron-feeded FTIR spectrometer at the SMIS beamline at SOLEIL synchrotron ( the high-flux broadband IR source is essential for measuring good signal to noise ratio transmission spectra) with combinatory recording of Raman signal and visual observation.

Fig. 1 summarizes the main data obtained on the hydrogen sample in the T-DAC at 80 K. Three photographs (Fig.1a) show the changes of the sample appearance upon pressure increase and decrease. The transition to black hydrogen is reversibly observed here at 310 GPa , as previously reported[21]. At 427 GPa, it is much more harder to distinguish the hydrogen sample from the gasket. As the toroidal shape enables an optimum elastic deformation for the diamond anvils tip, a very discernible Raman diamond edge, used as the pressure gauge, could be measured up to the maximum pressure and on release. The raw IR intensity transmission spectra were collected in the 800 cm$^{-1}$ – 8000 cm$^{-1}$ range (Fig. 1b). Up to 300 GPa, the signal intensity decreased due to change of the hydrogen sample size and of the deformation of the toroidal tip[8]. The strong absorption peak around 4000 cm$^{-1}$ is associated to the H2 vibron that appears above 160 GPa upon transiting from phase II to phase III, as reported previously[23]. Above 300 GPa, the sample configuration remained unchanged, therefore the changes of the IR spectra are only due to intrinsic absorption effects in hydrogen ( intensity at 300 GPa was recovered after pressure release). The H2 vibron mode broadens and shifts to lower wavenumbers with increasing pressure. Above 350 GPa, the intensity of the IR spectra is zeroed at high wavenumbers due to the progressive closure of the hydrogen solid electronic band gap below 1 eV.

Absorbance spectra have been obtained by ratioing the spectra with the reference spectrum taken at 123 GPa ( after intensity normalization with the 300 GPa spectrum). Full absorption is observed (absorbance value = 2), and progressing gradually towards lower wavenumbers with increasing pressure. Since the hydrogen sample was about 1.5 µm thick, the value of the hydrogen absorption coefficient corresponding to the full absorption plateau is estimated about 30 000 cm$^{-1}$, a typical value for a direct electronic bandgap. The direct bandgap energy is positioned at the junction between the absorbance plateau and the lower energy tail (the value of the excitonic level and direct bandgap should be almost identical[24]). The observed absorbance spectra can be explained by the combined effect of a direct bandgap and an indirect band gap with a slightly lower energy. That is compatible with the electronic band structures[24] of C2/c-24, calculated the most stable phase in this pressure range[25]. In figure 2 b, the direct electronic bandgap is seen to decrease linearly with pressure, also being well on line with the previous measurements in the visible[21]. After reaching a value of 0.5 eV at 425 GPa, a discontinuous evolution is observed, as evidenced by the complete absorption of the IR intensity over the whole spectral range. In figure 2c&d, the discontinuity is clearly manifested by looking at the transmitted IR intensity over the 800 cm$^{-1}$ - 2000 cm$^{-1}$ wavenumber range and at the evolution of its integral value with pressure. The intensity in this low wavenumber range is observed to discontinuously go to zero. That rules out a gradual closure of an indirect gap and the formation of a semi-metal with a plasma frequency progressing from below 0.1 eV with pressure. It is interesting to note that at the metal transition there is a confluence of the value of the bandgap and of the vibron energy. Upon pressure release, the infra-red transmission is discontinuously recovered and the pressure evolution of the IR absorbance of the hydrogen reversibly measured with pressure ( see Extended data figure 5). One can thus conclude that, at 425 GPa, a metal is formed with a plasma frequency greater than 1 eV. In calculations, the band gap is seen to be profoundly affected by the level of description of electronic exchange- correlation (XC) and also by NQEs. The local XC density functional theory calculation , such as PBE, which has been extensively used to search for stable structures in high pressure hydrogen[25], gives a gap that closes below 400 GPa. Including NQEs should lower this gap value by at least 1 eV[26]. As seen in fig.2c, the more advanced methods of quasi-particle approaches, GW[24] and DMC[9], give higher energy bandgap values than experiment but, taking into account NQEs, should be reduced. The use of an elaborate nonlocal functional, such as vdW-DF2, and including NQEs bring a reasonable agreement with experiment[26].

In phase I, the molecules are in a quantum free rotational state and arranged on a hexagonal close-packed lattice. Upon transiting to phase II, very small discontinuities in the lattice parameters have been measured[27]. Quantum molecular rotations become restricted and phase II can be described as a Quantum Fluxional Solid (QFS)[28]. At 160 GPa, phase II transforms into an ordered molecular phase III. High pressure structures of hydrogen have been extensively investigated using various levels of electronic calculations[4]. The best candidate for phase III, the C2/c-24 structure, was first discovered using an ab-initio random structure searching method, using the DFT-PBE[25]. It consists of layers of molecules whose bonds lie within the planes of the layers, forming a slight monoclinic distortion of the hexagonal lattice. The C2/c-24 has the lowest enthalpy and displays an intense IR vibron. It should also have a single intense lattice IR active phonon mode, which is reported here for the first time (Fig. 3a). The vibron frequency shift with pressure was reversibly observed upon pressure decrease (see Extended data Fig. 6).As reported in fig.3a, the calculated pressure evolutions of C2/c-24 vibron and phonon IR frequencies are in very good agreement with experimental data[25,29]. The phonon mode could be followed only to 225 GPa, above which it becomes hidden by the strong absorption of the diamond anvils. The linear vibron shift with pressure from 160 GPa to 425 GPa indicates that no structural change occurs up to 425 GPa. Therefore, the C2/c-24 structure is stable up to 425 GPa above which a first order transition to metal hydrogen occurs. However, in the framework of DFT-PBE, the C2/c-24 should be stable only up to 270 GPa[25]. Using DMC simulation and including NQEs, a transition from insulator C2/c-24 to metallic Cmca-12 was obtained at 424 GPa[9], associated to ordering of the molecular axis flat in the layers and to a larger monoclinic distortion of the hexagonal layers. This calculated pressure in excellent agreement with the observed transition to metal hydrogen. Furthermore, the fact that the transition is reversible with almost no hysteresis in pressure is a strong indication that the insulator to metal transition takes place within the molecular crystal. Following the DMC calculation, the atomic metal should be observed above 447 GPa, but its characterization is beyond the capability of the infrared approach used. An updated phase diagram of low temperature solid hydrogen is shown in fig.3b.

Searching for the intriguing properties of metal hydrogen will continue to foster novel high pressure approaches and the quality of the hydrogen sample obtained here above 400 GPa using T-DAC will ease measurements. As such, a non-invasive IR reflectivity measurement has recently been proposed[30] to characterize the predicted high temperature superconductivity both in the molecular and atomic hydrogen metal phases, respectively at 250 K[14] and 350 K[15]. Also, the next generation of synchrotron X-ray sources should enable to extend the structural characterization of solid hydrogen up the insulator-metal phase transition.

**Acknowledgements:**

We thank Olivier Marie for FIB machining the Toroidal anvils and the gasket hole. We are grateful to the direction of SOLEIL for continuously providing the infrared beamtime over the past 6 years. We thank Ferenc Borondics and Francesco Capitani for help on the SMIS beamline.

**Author contributions**:

P.L. designed the project. P.L and F.O. prepared and loaded the T-DAC. P.D, F.O and P.L developed the IR microscope. P.L, F.O and P.D conducted the experiment and analyzed the data. P.L and P.D wrote the manuscript. All authors discussed the results.

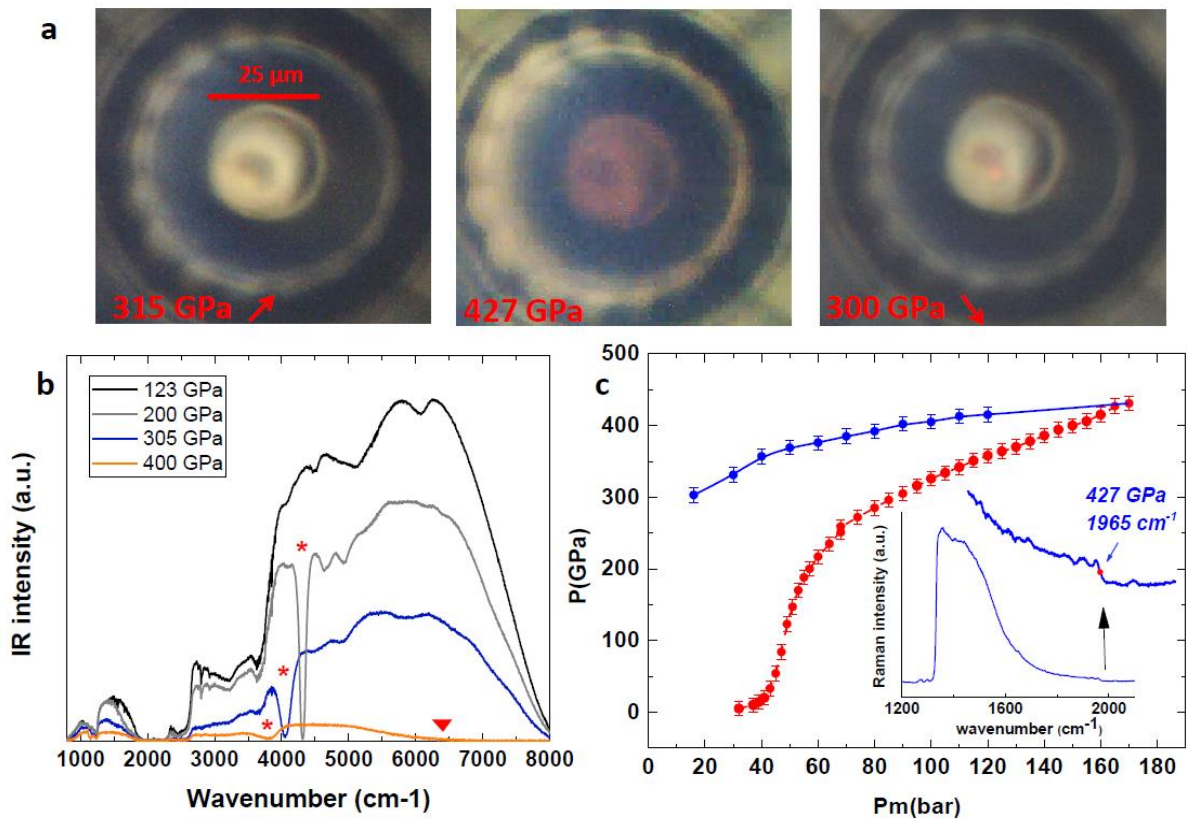

**Figure 1.** Measurements in hydrogen above 300 GPa. **a.** Photographs of the hydrogen sample taken at different stages of compression, with a bright light illumination also on the back side. At 310 GPa, the sample turns black and non-transmitting reversibly as illustrated by the photographs at 315 GPa on increasing pressure and at 300 GPa on decreasing pressure. At 427 GPa, the sample is in the metallic state and hardly seen from the Re gasket. The red color at the diamond tip center is attributed to the closure of the diamond band gap[8]. **b.** IR transmission spectra at selected pressures. Intrinsic absorption of the hydrogen sample, associated to the vibron and the closure of the electronic bandgap, are respectively indicated as stars and triangle. **c.** Pressure evolution in hydrogen versus the helium membrane pressure pushing the piston of the DAC, red and blue dots indicate pressure increase and decrease, respectively. Inset, the Raman diamond spectrum collected at 427 GPa: the high frequency edge, used as the pressure gauge, corresponds to the diamond layer at the interface with hydrogen and is indicated by a red dot.

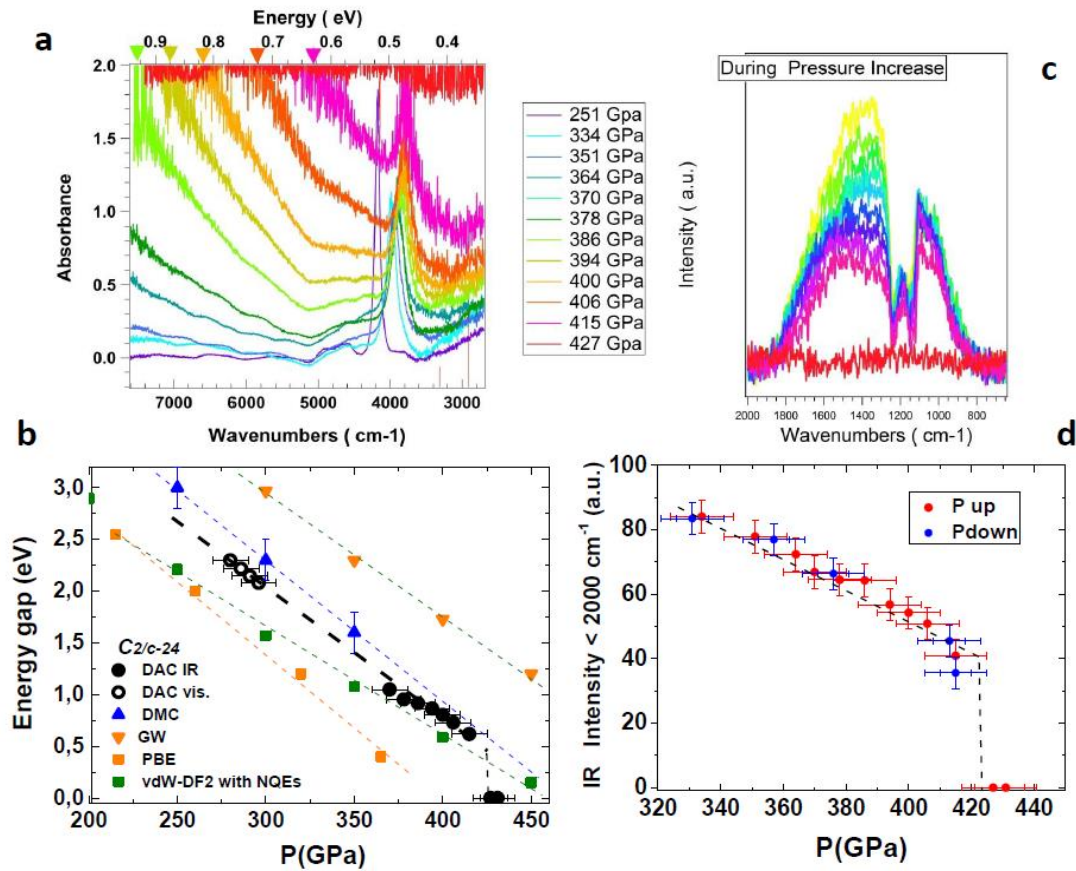

**Figure 2.** Closure of the direct electronic bandgap and first order phase transition to metal hydrogen. **a**. Absorption spectra of hydrogen at different pressures. Above 386 GPa the shape of the absorption edge enables to position the direct electronic bandgap energy, as triangles. Different colors are associated to different pressures for clarity. **b**. The pressure evolution of the experimental electronic bandgap, combining previous data in the visible[21] and present IR data, is compared to various calculations for the C2/c-24 structure performed under various approximations: within the DFT framework with local, PBE, and nonlocal, vdW-DF2, XC functionals [26]; with the quasiparticle approach within the GW approximation[9]; with DMC method[24]. **c**. Transmission spectra over the medium IR range, 800 cm$^{-1}$ – 2000 cm$^{-1}$. The pressure color scale is as in panel a. **d.** Transmitted intensity integrated over the medium IR range versus pressure. The red and blue dots indicate pressure increase and release, respectively.

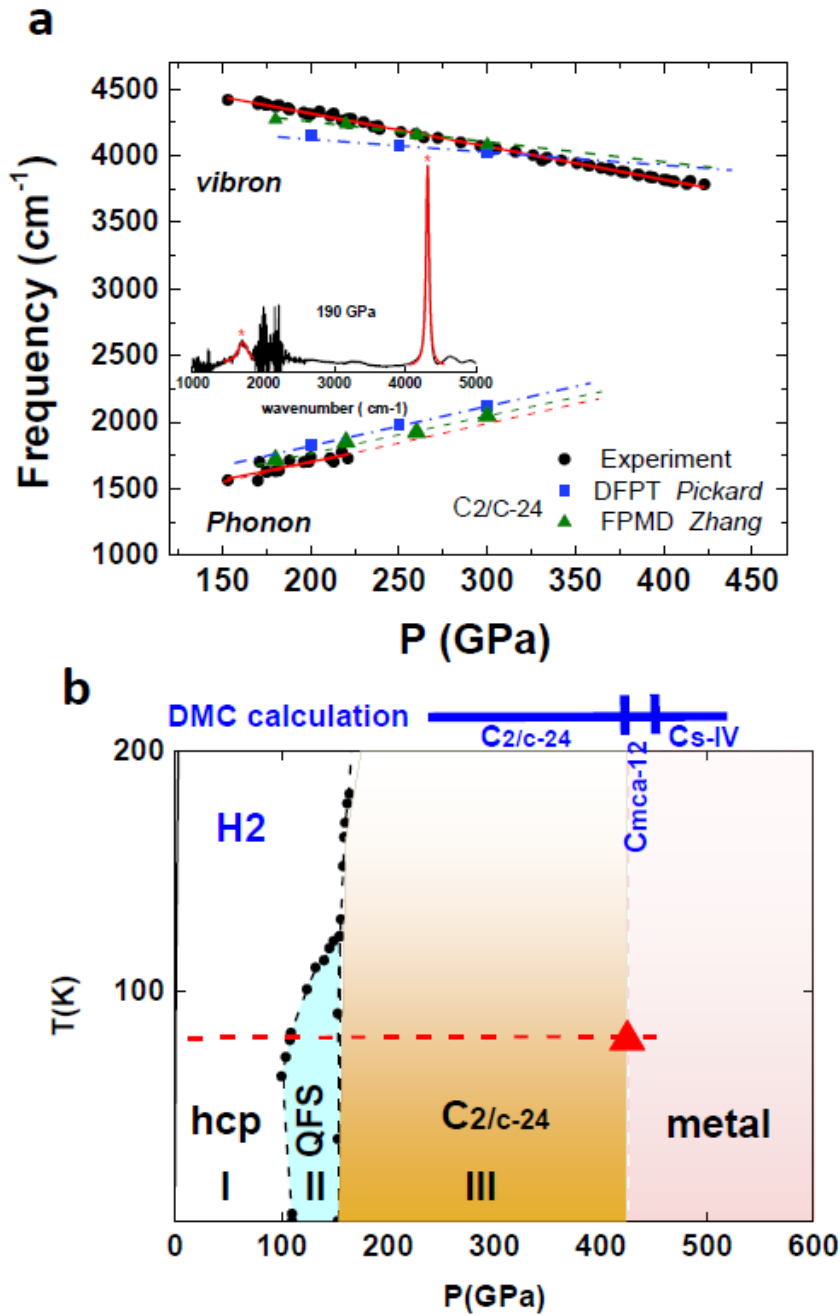

**Figure 3.** Evolution with pressure of solid hydrogen IR active modes and phase diagram of hydrogen at low temperature. **a**. Frequencies of the vibron and the phonon IR active modes of the C2/c-24 structure versus pressure. The black dots and red lines represent experimental data and linear fits. The blue square and green triangles are the density functional perturbation theory (DFPT)[25] and first-principles molecular dynamics (FPMD)[29] calculations. Inset: the IR absorption spectrum at 190 GPa showing the vibron and phonon peaks. **b**. Experimental phase diagram of hydrogen at low temperature. The red dash line is the pathway of the IR data collection and the red triangle the transition to metal hydrogen. Boundary lines between phases I, II and III are from previous studies[4]. Above the graph, the phase transitions sequence calculated by the Diffusion Monte Carlo including NQEs[9].

# Methods.

**The Toroidal Diamond Anvil Cell**. The Toroidal shape at the tip of the synthetic single-crystal diamond anvil was made by focused ion beam machining. Scanning electron micrograph picture and profile of the toroidal tip are given in Extended Data fig.1. The central flat and the groove are 25 μm and 80 μm in diameter, respectively. The toroidal tip has been recovered intact upon pressure release, indicating its purely elastic deformation up to the highest pressure. High pressures were generated using the LeToullec membrane DAC equipped with Boehler-Almax type seats made of polycrystalline diamond. The hydrogen sample was loaded in the DAC under 140 MPa. A rhenium gasket was used. The pressure was very smoothly increased by inflating the membrane with a slow rate of 0.2 bar/min to allow gradual elastic deformation of the anvil tip and also enabling a minimum misalignment of the sample in the cryostat. The conversion between the membrane pressure and the force on the piston is F(kN)= 0.05xPm(bar). A red color of the diamond tip is observed above 400 GPa, caused by the onset of an absorption of visible light at the diamond tip, and reversibly disappearing upon pressure release.

**Pressure measurement.** The high-frequency edge of the T2g Raman band of the diamond anvil at the hydrogen sample interface has been used to measure the pressure. The hydrogen sample pressure was related to the diamond spectra high frequency edge following the Akahama's calibration of 2006[31]. The revision of this calibration to go in the 400 GPa pressure range, proposed by Akahahama in 2010[32], was not selected because it seems to overestimate the pressure, based on the following facts: i) in our previous measurements using Toroidal anvil using X-ray diffraction8), the pressure obtained by the Re pressure gauge was in better agreement with 2006 Akahama's pressure scale. ii) As seen in the Extended data fig.2, the pressure evolution versus the membrane pressure and the IR vibron frequency versus pressure have unphysical behaviors if the 2010 Akahama scale is used instead of the 2006 Akahama one. The pressure of the band gap energy data measured previously in the visible21) has been corrected using the 2006 Akahama calibration. No difference in the diamond edge Raman pressure could be observed between the use of the toroidal anvil and normal anvil at least up to 330 GPa, by using the IR H2 vibron frequency versus pressure ( see Extended data fig. 6. The random error bars on the pressure determination is ±10 GPa.

**Infra-Red bench.** The photograph of the bench is shown in Extended Data Fig. 3. This bench was previously used to characterize the infrared vibrational modes of phase IV of hydrogen at 300K[33]. The custom-made horizontal infrared microscope is equipped with two infinity-corrected long working distance Schwarzschild's objectives ( 47 mm working distance, NA 0.5) that produce a 22 microns (FWHM) IR spot. The spatial and temporal stability of the broadband infrared beam allowed to record transmission spectra with less than 6% noise value through a 3 microns diameter hole (see extended data Fig.3). One of the Schwarzschild is mounted on a translation stage to free space behind the cryostat and to insert the optical head for Raman spectroscopy measurements without moving the DAC. Raman signal was excited by a 660 nm wavelength laser limited to 3 mW power output above 300 GPa to prevent thermal heating and breaking of the toroidal tip. The Raman head is also equipped with a digital camera to take photographs of the sample transformation under pressure increase. The quality of the measurements obtained also relies on the high mechanical stability of the bench and the high reproducibility in position upon sliding between the IR to the Raman configurations. Infrared spectra were collected with a 4 cm$^{-1}$ resolution and 1024 scans.

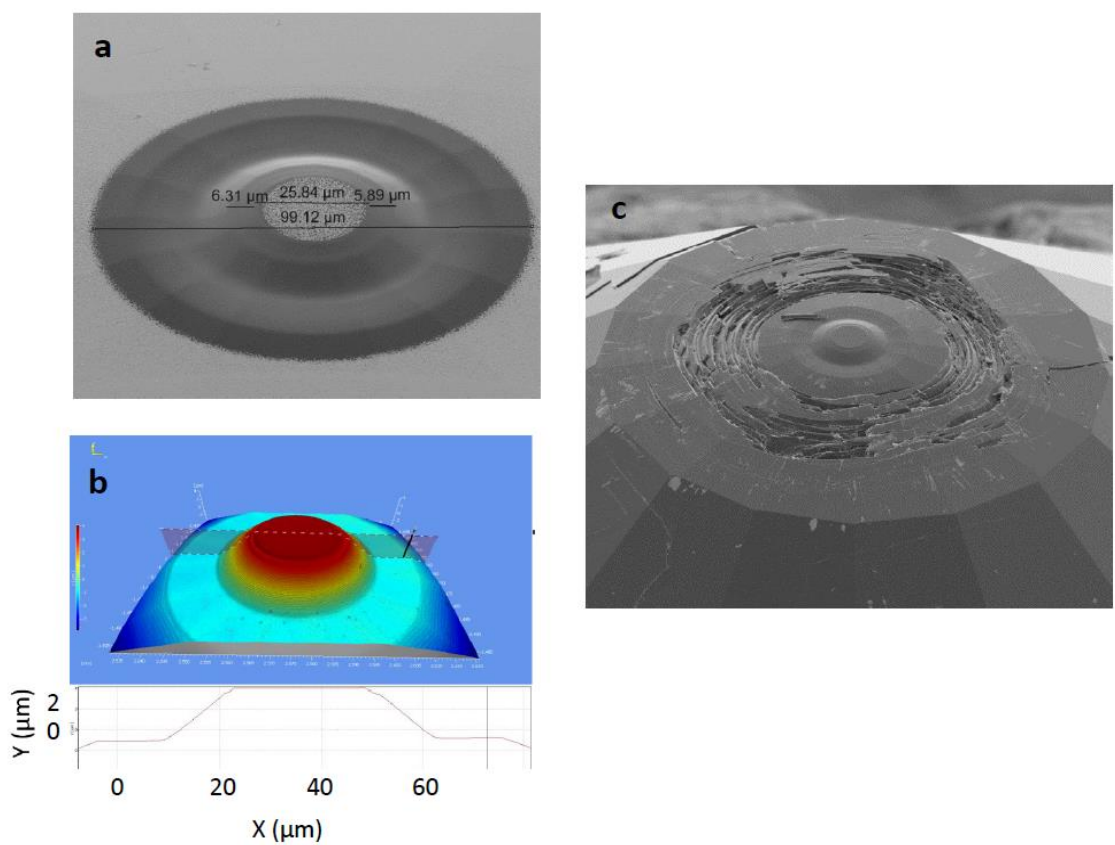

**Extended Data Figure 1 | Geometry of the Toroidal shape used**. **a**. Scanning electron micrograph picture of the anvil. **b**. Profile of the toroidal diamond tip measured by interferometry. **c**. Scanning electron picture of the toroidal anvil recovered after pressure release. The toroidal part of the anvil is intact. Ring cracks are seen on the bevel of the anvil, about the 150 microns diameter.

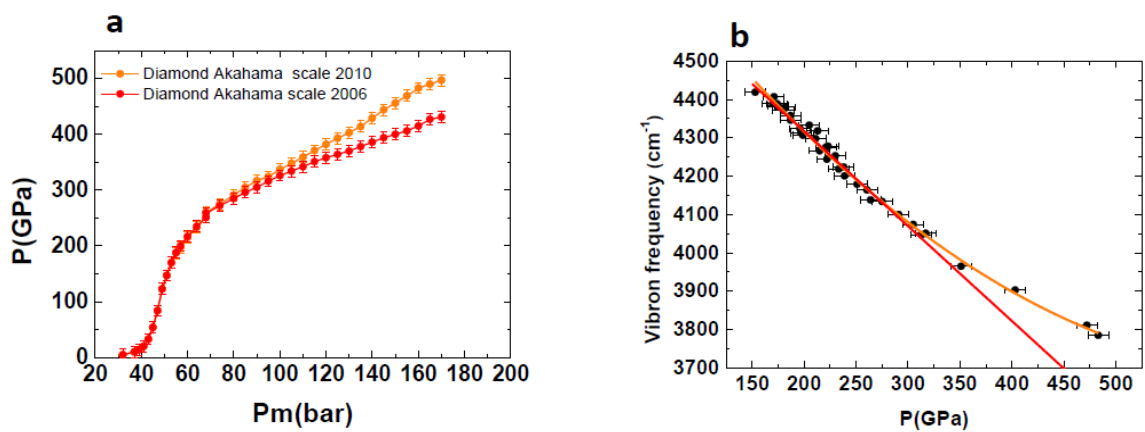

**Extended Data Figure 2 | Comparison of the Akahama 2006 and the Akahama 2010 diamond pressure scales on two measurements.** The two scales deviate above 300 GPa. **a**. The sample pressure versus the helium membrane pressure exerting the force on the piston of the DAC ( F(kN) = 0.05 x Pm(bar)). The 2010 scale[32] (orange) gives a convex evolution above 300 GPa that is not physical. **b**. The IR vibron frequency versus pressure. Above 300 GPa, the 2010 scale ( orange) gives a sublinear red shift of the frequency whereas the one corresponding to the 2006 scale[31] is linear in better agreement with calculations, taking into account the stability of the C2/c-24 phase.

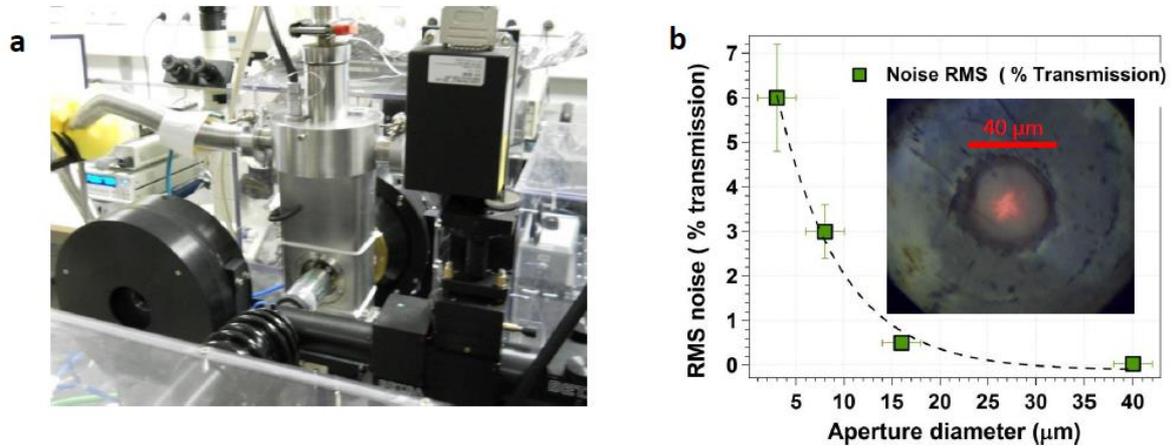

**Extended Data Figure 3 | Experimental setup on the horizontal infrared microscope at the SMIS beamline of the SOLEIL synchrotron**. **a**. The He-flow cryostat containing the T-DAC sits in between the two Scharzchild objectives for IR transmission measurements. In the Raman configuration, the optical transfer of the Raman setup is positioned in place of the collection Scharzchild objective. The Raman head and the Scharzchild objective are mounted on a long travel translation stage, enabling concomitant move and excellent reproducibility in positioning. **b**. RMS noise value of the transmission spectra recorded at 4 cm$^{-1}$ resolution and after 400 accumulations as a function of the gasket hole diameter. Inset: Photograph of the IR synchrotron beam focused in a 40 μm gasket hole in the DAC.

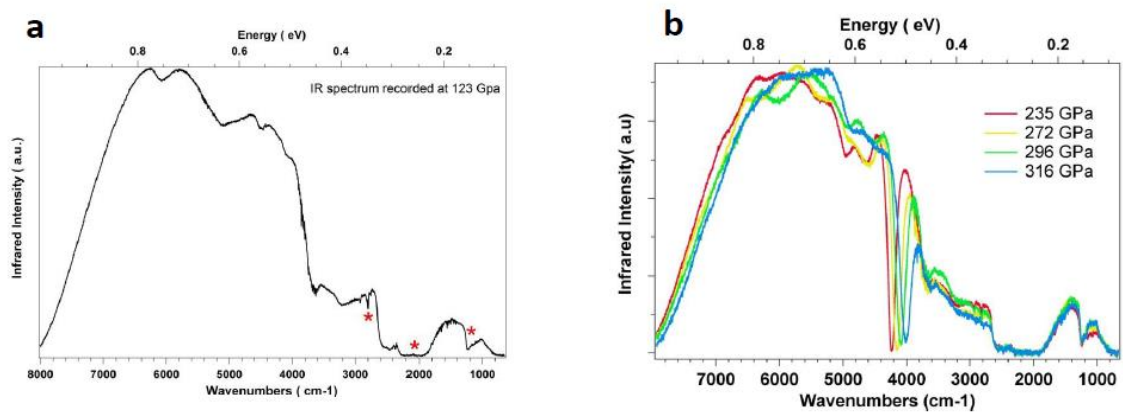

**Extended Data Figure 4 | IR transmission spectra below 300 GPa. a**. The single beam spectrum at 123 GPa which is used as a reference spectrum for absorption spectra ( after normalizing its intensity with the one at 300 GPa). The three red stars indicate parasitic effects corresponding to: absorption peaks of impurities around 2800 cm$^{-1}$; a broad absorption band from the diamond ( 1900 – 2300 cm$^{-1}$); a broad absorption around 1200 cm$^{-1}$ which comes from the protected layers of aluminum mirrors in the beamline. **b**. Single beam spectra, normalized to their signal intensity (Peak-peak value of their respective interferograms).

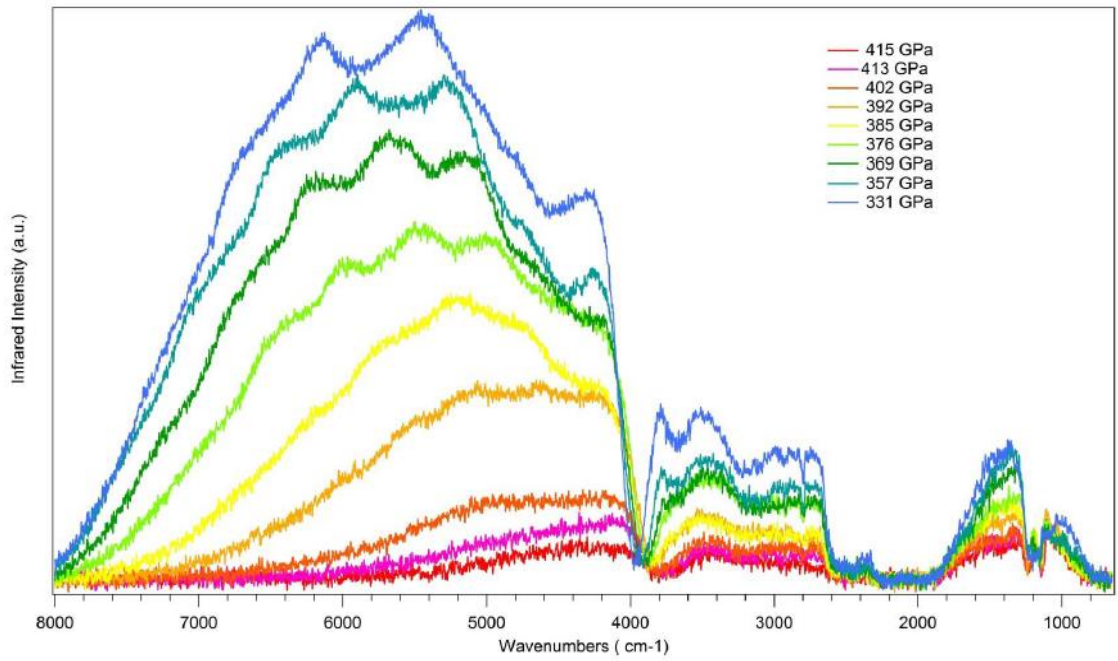

**Extended Data Figure 5 | IR transmission spectra at different pressures upon pressure release**.

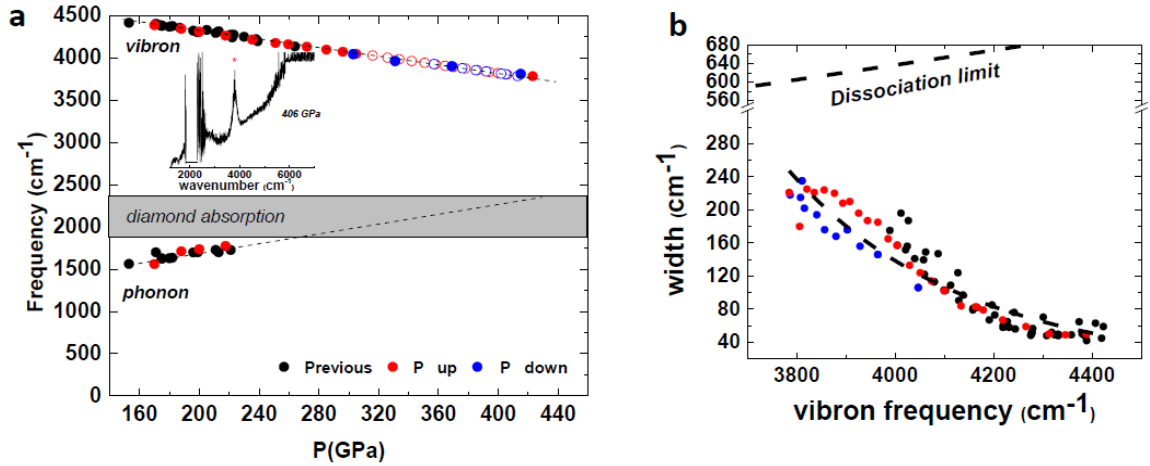

**Extended Data Figure 6 | IR Vibron and phonon frequencies versus pressure**. **a**. Various symbols correspond to: red, blue and black colors respectively to pressure increase, pressure decrease and previous campaign using standard DACs ; full dots, pressure measured by the diamond pressure scale and open symbols, positioned following the linear vibron shift with pressure. Inset IR absorption spectra at 406 GPa with the vibron peak indicated by a red star.

The linear fit of the vibron frequency with pressure is: $\upsilon_{vibron}$ (cm$^{-1}$) = 4814 -2.48 P(GPa).

The linear fit of the phonon frequency with pressure is: $\upsilon_{phonon}$ (cm$^{-1}$) =1163 +2.69 P(GPa).
**b.** Full Width, W, of the vibron peak versus its energy, E. The increase of the vibron linewidth with pressure remains much less than the dissociation value, given by W= E/2π. This limit is estimated by equating the lifetime of the vibrational state (inferred by assuming it is the only contribution to the line broadening) to the vibration period. Therefore, there is no sign of dissociation in the C2/c-24 phase prior to the transition to metal hydrogen.

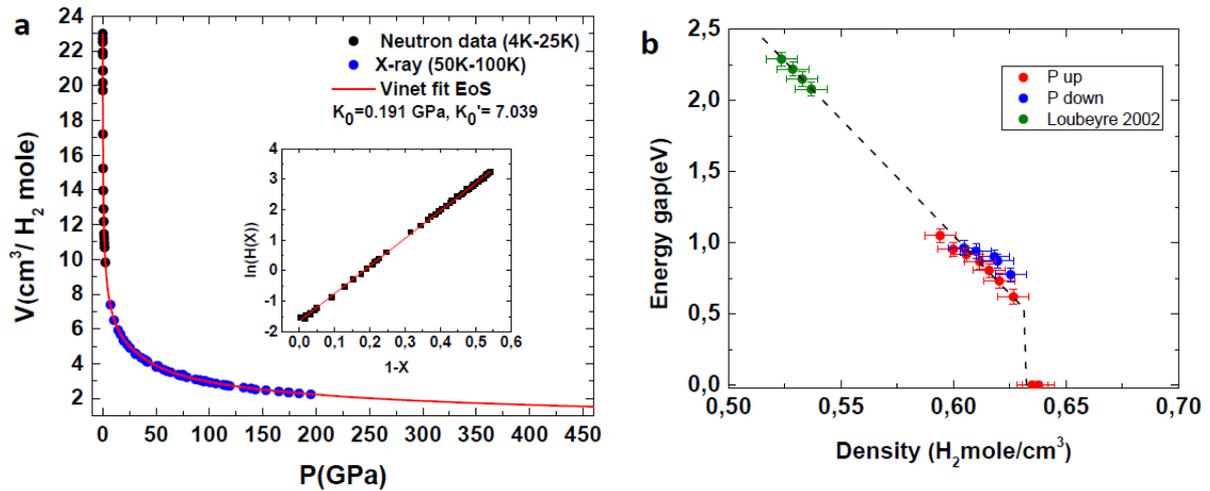

**Extended Data Figure 7 | Equation of state of hydrogen and evolution of the direct electronic bandgap versus density**. **a**. Equation of state of solid hydrogen around 80 K. The black dots are neutron diffraction measurements[34]. The blue dots are our unpublished x-ray diffraction data obtained in front of the ESRF synchrotron ( The pressure scale is the based on the revised ruby scale of 2008[35]). The red line is the fit of the experimental data by a Vinet form[36]: $P=3K_0(1-X)X^{-2}\exp(3/2(K_0'-1)(1-X))$, with $X=(V/V_0)^{1/3}$, $K_0=0.191$ GPa and $K_0'=7.039$. Inset: The Vinet form can be reformulated in terms of expressions analogous to normalized stress, $\ln(H(X)) = \ln(PX^2/3(1-X))$, and Eulerian strain, $(1-X)$. This gives: $\ln(H(X) = \ln K_0 + 3/2(K_0'-1)(1-X)$. The linear fit of the data is shown. The present EoS is in good agreement with the one measured by Akahama et al[37]. **b**. Evolution of the electronic direct bandgap with density. The color of the data points corresponds to: red, IR data pressure increase; blue, IR data pressure decrease; green, data in the visible[21].

a

| Vibron H2 (cm⁻¹) | Energy gap(eV) | P (GPa) |
| --- | --- | --- |
| 3894 | 1.05 | 370 |
| 3877 | 0.95 | 378 |
| 3856 | 0.92 | 386 |
| 3836 | 0.87 | 394 |
| 3821 | 0.80 | 400 |
| 3806 | 0.73 | 406 |
| 3785 | 0.62 | 415 |
|  | 0 | 427 |

b

| Vibron H2 (cm⁻¹) | Energy gap(eV) | P (GPa) |
| --- | --- | --- |
| 3787 | 0.77 | 413 |
| 3808 | 0.87 | 405 |
| 3816 | 0.90 | 402 |
| 3841 | 0.94 | 392 |
| 3857 | 0.96 | 385 |

**Extended Data table 1 | IR vibron frequency and direct electronic bandgap.** The pressure uncertainty is ±10 GPa; the gap energy uncertainty ±0.05 eV, the vibron frequency uncertainty ± 4 cm⁻¹.